\documentclass[conference]{IEEEtran}
\usepackage{spconf,amsmath,graphicx,subfigure}
\usepackage{booktabs}

\usepackage{multirow}
\usepackage{array}

\title{Endoscopy disease detection challenge 2020}

\name{%
\begin{tabular}{@{}c@{}}
Sharib Ali$^{1}$ \qquad 
Noha Ghatwary$^{7}$ \qquad 
Barbara Braden$^{2}$ \qquad 
Dominique Lamarque$^{3}$ \qquad \\
Adam Bailey $^{2}$\qquad 
Stefano Realdon$^{4}$ \qquad 
Renato Cannizzaro $^{5}$  \qquad 
Jens Rittscher$^{1}$  \qquad\\
Christian Daul$^{6}$  \qquad 
James East $^{2}$
\end{tabular}}

\address{$^{1}$ Institute of Biomedical Engineering, Big Data Institute, University
of Oxford, \\Old Road Campus,  Oxford, UK\\
$^{2}$ Translational Gastroenterology Unit, Experimental Medicine Div., \\ John Radcliffe Hospital, University of Oxford, Oxford, UK \\
$^{3}$ Universit{\'e} de Versailles St-Quentin en Yvelines, H{\^o}pital Ambroise Par{\'e}, France\\
$^{4}$ Instituto Onclologico Veneto, IOV-IRCCS, Padova, Italy\\
$^{5}$ CRO Centro Riferimento Oncologico IRCCS Aviano Italy\\
$^{6}$ CRAN UMR 7039, University of Lorraine, CNRS, Nancy, France\\
$^{7}$ University of Lincoln, UK\\
}
\begin{document}
\maketitle
\thispagestyle{empty}
\pagestyle{empty}
\begin{abstract}
Whilst many technologies are built around endoscopy, there is a need to have a comprehensive dataset collected from multiple centers to address the generalization issues with most deep learning frameworks. What could be more important than disease detection and localization? Through our extensive network of clinical and computational experts, we have collected, curated and annotated gastrointestinal endoscopy video frames. We have released this dataset and have launched disease detection and segmentation challenge EDD2020\footnote{https://edd2020.grand-challenge.org} to address the limitations and explore new directions. EDD2020 is a crowd sourcing initiative to test the feasibility of recent deep learning methods and to promote research for building robust technologies. In this paper, we provide an overview of the EDD2020 dataset, challenge tasks, evaluation strategies and a short summary of results on test data. A detailed paper will be drafted after the challenge workshop with more detailed analysis of the results.
\end{abstract}
\section{INTRODUCTION}
Endoscopy is a widely used clinical procedure for the early detection of cancers in hollow-organs such as oesophagus, stomach, colon and bladder. During this procedure an endoscope is used; a long, thin, rigid or flexible tube with a light source and camera at the tip to visualize the inside of affected organs on an external screen. Computer-assisted methods for accurate and temporally consistent localization and segmentation of diseased region-of-interests enable precise quantification and mapping of lesions from clinical endoscopy videos which is critical for monitoring and surgical planning. Innovations have the potential to improve current medical practices and refine health-care systems worldwide. However, well-annotated, representative publicly available datasets for disease detection for assessing research reproducibility and facilitating standardized comparison of methods is still lacking. Many methods to detect diseased regions in endoscopy have been proposed however these have been primarily focused on the task of polyp detection in the gastrointestinal tract with demonstration on datasets acquired from at most a few data centers and single modality imaging, most commonly white light. Here, we present our multi-class, multi-organ and multi-population disease detection and segmentation challenge in clinical endoscopy. With this sub-challenge we aim to establish a comprehensive dataset and benchmark existing and novel algorithms for disease detection.

Specifically, we aim to assess localization of disease regions using bounding boxes and exact pixel-level segmentation. Clinical applicability by assessing real-time monitoring and offline performance evaluations of algorithms for improved accuracy and better quantitative reporting is required today. In this challenge, participants are provided with 380 annotated video frames by medical experts and experienced post doctoral researchers. The dataset is comprehensive and consists of endoscopy frames from different gastrointestinal tract organs and varied endoscopy modalities. It incorporates multiple populations with 4 different international centers and include below mentioned categories:
\begin{itemize}
\item Organ 1: Colon, associated disease: Polyp, cancer
\item Organ 2: Oesophagus, associated disease: Barrett’s, dysplasia and cancer
\item Organ 3: Stomach, associated disease: Pyloric inflammation, dysplasia
\end{itemize}

\section{Proposed tasks}
To rigorously assess the methods for disease detection and localization in various organs, participants are requested to complete two sub-tasks on the provided dataset:
\begin{itemize}
\item[1.] Disease detection and localization: This task is evaluated based on the inference results on the test dataset provided from a subset of the data collected for the training. 
\item[2.] Disease area segmentation: This task will be evaluated based on the results of the test dataset provided from a subset of the data collected for training
\item[3.] Out-of-sample detection: This task will be evaluated on a completely unknown set of images which have not been used in either of training or testing datasets for other tasks
\end{itemize}
Participants are required to identify, localize and segment five classes consisting of: 1) Normal dysplastic Barrett's Oesophagus (NDBE), 2) suspicious area (``subtle pre-cancerous lesion''), 3) high-grade dysplasia, 4) adenocarcinoma (``cancer''), and  5) polyp. Two or more classes can be overlapped, for e.g., Barrett's can colocalize with low-grade or high-grade dysplastic area. 
\section{Dataset}
The EDD2020 dataset \footnote{http://dx.doi.org/10.21227/f8xg-wb80} is unique and consists of five classes for both detection and segmentation tasks. All data are from the gastrointestinal tract that comprise of oesophagus, stomach, colon and rectum. The data is collected from Ambroise Par{\'e}Hospital of Boulogne-Billancourt, Paris, France; Centro Riferimento Oncologico IRCCS, Aviano, Italy; Istituto Oncologico Veneto, Padova, Italy and John Radcliffe Hospital, Oxford, UK. All the collected data were annonymised at pre-collection stage by the primary institution and by the EDD2020 challenge organisers before releasing the data publicly. No patient information has been provided or communicated in this process and all data has been acquired through patient consent at primary host institutions.

\begin{figure*}
\centering
\includegraphics[scale=0.7]{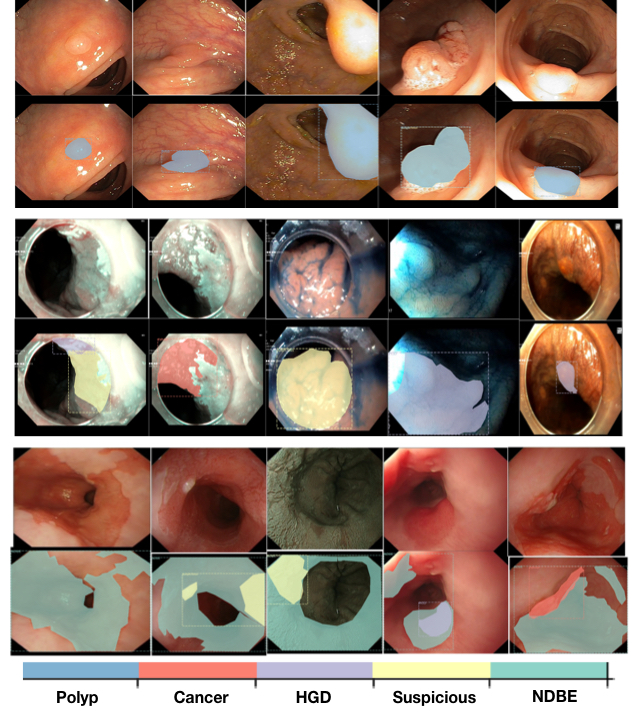}
\caption{Endoscopy disease detection and segmentation dataset (EDD2020). The top two rows consists of mostly polyps, middle two rows consists of mostly suspected and high-grade dysplastic (HGD) cases in stomach, and finally bottom two row represents non-dysplastic Barrett's oesophagus (NDBE), suspected dysplasia and one case for cancer.  \label{fig:EDD2020}}
\end{figure*}

Annotations were performed by two clinical experts and two post-doctoral researchers with a background in endoscopy imaging. Due to annotation variability and difficult to get consensus between experts in some cases of this dataset we would like to declare that these annotations are subjective to the annotators involved in the preparation of this dataset. There were four different iterations of this dataset. The dataset consists of 386 still image frames acquired from multiple videos.  The ground truth for training data consists of 160 manually labeled annotations for non-dysplastic Barrett's, 88 cases for suspicious precancerous lesion,  74 for high-grade dysplasia,  53 cancer and 127 polyp masks with a total of 503 ground truth annotations. The same number of bounding boxes also exists for each of these cases. Figure~\ref{fig:EDD2020} shows label annotations for some frames in EDD2020 dataset. The bounding box annotations are provided in \textit{PASCAL VOC} format~\cite{pascal-voc-2012} and the label masks are provided as 5 channel \textit{tif} images. An open-source VIA annotation tool~\cite{dutta2019vgg} was used for all annotations.

The test dataset is restricted to be used only for this challenge and is made available to the participants of this challenge only. Methods are evaluated on an online leaderboard and will be analysed at the end of the competition. Codes to help participants getting started have been made available online\footnote{https://github.com/sharibox/EDD2020}.
\section{Evaluation}
The challenge problems fall into three distinct categories. For each there exists already well-defined evaluation metrics used by the wider imaging community which we use for evaluation here. Codes related to all evaluation metrics used in this challenge are also available online\footnote{https://github.com/sharibox/EndoCV2020}.
\subsection{Detection score}
Metrics used for multi-class disease detection: 
\begin{itemize}
    \item IoU - intersection over union. This metric measures the overlap between two bounding boxes $A$ and $B$ as the ratio between the overlapped area $A\cap B$ over the total area $A\cup B$ occupied by the two boxes:
    \begin{equation}
        \mathrm{IoU} = \frac{A\cap B}{A\cup B}
    \end{equation}
    where $\cap$, $\cup$ denote the intersection and union respectively. In terms of numbers of true positives (TP), false positives (FP) and false negatives (FN), IoU (aka Jaccard J) can be defined as:
    \begin{equation}
        IoU/J = \frac{TP}{TP+FP+FN}
    \end{equation}
    
    \item mAP - mean average precision of detected disease class with precision (p) defined as $p={\frac {TP}{TP+FP}}$ and recall (r) as $r={\frac {TP}{TP+FN}}$. {This} metric measures the ability of an object detector to accurately retrieve all instances of the ground truth bounding boxes. The higher the mAP the better the performance. Average precision (AP) is computed as the Area Under Curve (AUC) of the precision-recall curve of detection sampled at all unique recall values $(r_1, r_2,...)$ whenever the maximum precision value drops:
    \begin{equation}
        \mathrm{AP} = \sum_n{\left\{\left(r_{n+1}-r_{n}\right)p_{\mathrm{interp}}(r_{n+1})\right\}}, 
    \end{equation}

    with $p_{\mathrm{interp}}(r_{n+1}) =\underset{\tilde{r}\ge r_{n+1}}{\max}p(\tilde{r})$. Here, $p(r_n)$ denotes the precision value at a given recall value. This definition ensures monotonically decreasing precision. The mAP is the mean of AP over all disease classes $i$ for $N=5$ classes given as
    \begin{equation}
        \mathrm{mAP} = \frac{1}{N}\sum_i{\mathrm{AP}_i}
    \end{equation}
    This definition was popularised in the PASCAL VOC challenge~\cite{pascal-voc-2012}. 
    The final mAP (mAP$_d$) was computed as an average mAPs for IoU from 0.25 to 0.75 with a step-size of 0.05 which means an average over 11 IoU levels are used for 5 categories in the competition (mAP $@[.25:.05:.75]$ ).
\end{itemize}
Participants were finally ranked on a final mean score $(\mathrm{score_d})$, a weighted score of mAP and IoU represented as (for details please refer to \cite{ALI2020:SREP}):
\begin{equation}
\mathrm{score_d} = 0.6 \times \mathrm{mAP_d} + 0.4 \times \mathrm{IoU_d}
\end{equation}

Standard deviation between the computed mAPs ($\pm \sigma_{score_d}$) are taken into account when the participants have the same $\mathrm{score_d}$. 
\subsection{Semantic segmentation score}
Metrics widely used for multi-class semantic segmentation of disease classes have been used for scoring semantic segmentation. It comprises of an average score of $F_1$-score (Dice Coefficient), $F_2$-score, precision and recall.

\textbf{Precision, recall, F$_\beta$-scores:}
These measures evaluate the fraction of correctly predicted instances. Given a number of true instances $\#\mathrm{GT}$ (ground-truth bounding boxes or pixels in image segmentation) and number of predicted instances $\#\mathrm{Pred}$ by a method, precision is the fraction of predicted instances that were correctly found, $p=\frac{\#\mathrm{TP}}{\#\mathrm{Pred.}}$ where $\#\mathrm{TP}$ denotes number of true positives and recall is the fraction of ground-truth instances that were correctly predicted, $r=\frac{\#\mathrm{TP}}{\#\mathrm{GT}}$. Ideally, the best methods should have jointly high precision and recall. $\mathrm{F_\beta}$-scores gives a single score to capture this desirability through a weighted ($\beta$) harmonic means of precision and recall, $\mathrm{F_\beta} = (1 + \beta^2) \cdot \frac{{p} \cdot {r}}{(\beta^2 \cdot {p}) + {r}}$.
  
 Participants are ranked on as the increasing value of their semantic performance score given by:
\begin{equation}
\mathrm{score_s} = 0.25* \sum p + r + \mathrm{F_1} + \mathrm{F_2} 
\end{equation}
Standard deviation between of each subscores are computed and averaged to obtain the final $\pm \sigma_{score_s}$ which is used during evaluation for participants with same final semantic segmentation score.
\subsection{Out-of-sample detection score}
Out-of-sample generalization of disease detection is defined as the ability of an algorithm to achieve similar performance when applied to a completely different institution data. We do not provide any details of this data in this paper as this will be released as a part of test data during the competition. To assess this, participants should apply their trained models on images that were neither included in the training nor test data of the detection and segmentation tasks. Participants will be ranked based on a score gap of generalization defined as:
\begin{equation}
    \mathrm{dev_g} =\mid \mathrm{mAP_d} - \mathrm{mAP_g}\mid
\end{equation}
It is worth noting that the highest $\mathrm{mAP_g}$ with $\mathrm{dev_g} \rightarrow 0$ is required for winning the competition.

{The deviation score $\mathrm{dev_g}$ can be either positive or negative, however, the absolute difference should be very small (ideally, $\mathrm{dev_g} = 0$). The best algorithm should have high $\mathrm{mAP_g}$ and high $\mathrm{mAP_d}$ but a very low $\mathrm{dev_g} (\rightarrow 0$). 

%
\section{Analysis on EDD2020 test data}
It can be observed in Fig. ~\ref{fig:EDD2020_algorithms} (A) that all detections of the test classes and bounding box localisation are optimal and accurate for algorithm-I. It is to be noted that most papers dealing with endoscopy based detection only use classical metrics according to which here the classification accuracy is 100\% for the above samples. However, we discourage such concept and use well-established computer vision metrics to compute the average mAP for different IoU thresholds~\cite{pascal-voc-2012}. It is to be noted that such metrics showing an accuracy over 35\% mAP is already high enough for clinical translation provided its real-time processing time and robustness to patient variability. The average mAP for 8 images shown is 58.52\% with mAP$_{25}$ and mAP$_{50}$ of 62.5\% and 62.5\%, respectively.

However, the overall mAP on the entire dataset is just 35\% which is mostly because of not enough samples of adenocarcinoma samples and samples of polyp mixed with some Barrett's topology as shown in  Fig. ~\ref{fig:EDD2020_algorithms} (B). We have used another baseline method 2 and method 3 to correct for these wrong labels. An ensemble result provided the best score over 45\% on the average mAP. Detailed methods will be written as a journal submission in future. 
\begin{figure*}
\centering
\includegraphics[scale=0.7]{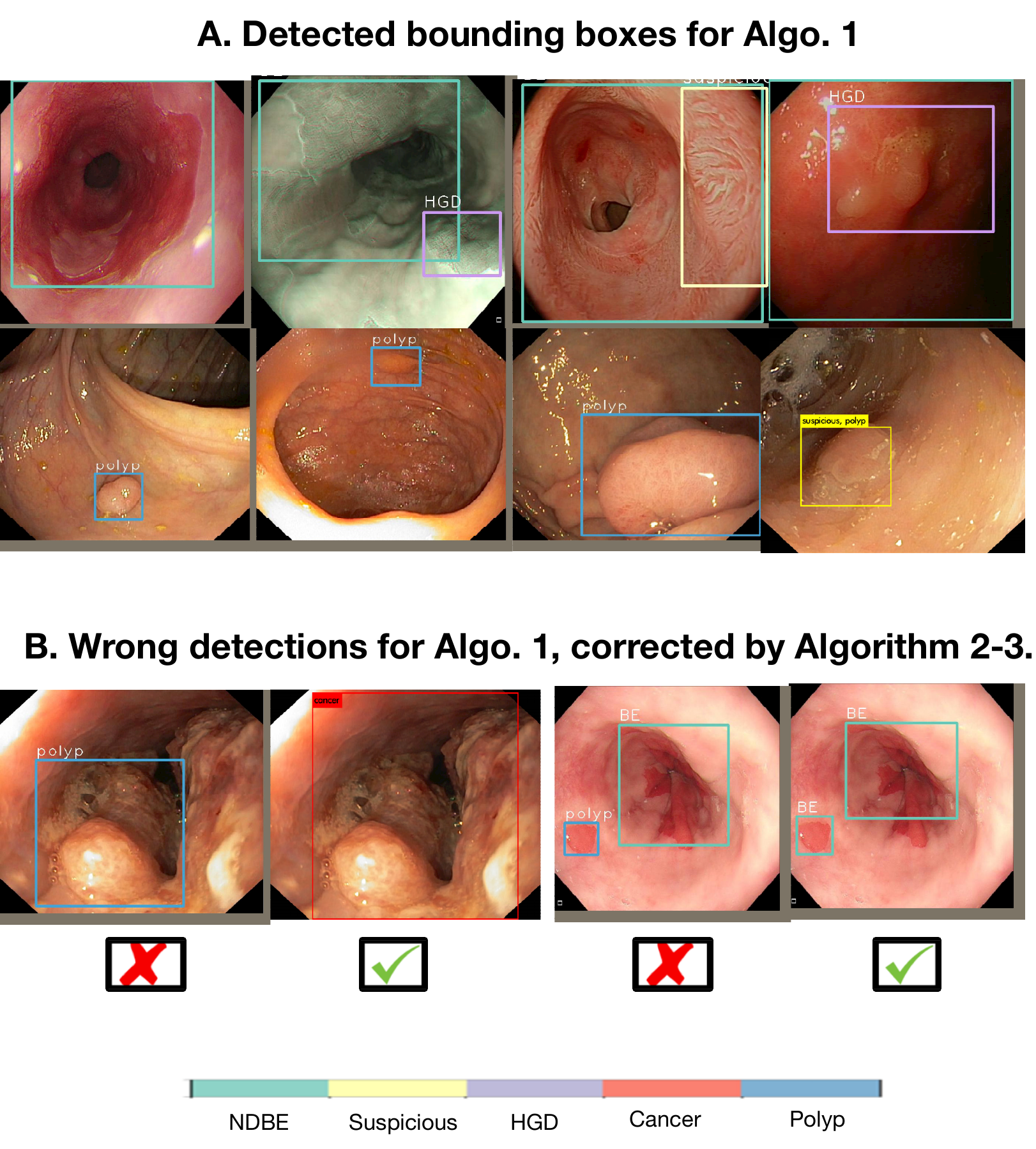}
\caption{Detectection and localisation of disease classes on EDD2020  test dataset. A) Consists of correctly detected bounding boxes for all cases in the test data using our in-house Algorithm 1. It can be observed that The algorithm detects precise bounding boxes for non-dysplastic Barrett's (NDBE), high-grade and suspected (dysplasia). In the 2nd row, different polyps of various size and locations are localised.  B) Algorithm 1 fails to detect cancer instead it confuses with polyp like protrusion. Similarly, a small Barrett's island is confused with polyp. However, this is corrected by algorithm 2 and 3. \label{fig:EDD2020_algorithms}}
\end{figure*}
\section{Conclusion}
A comprehensive dataset using multiple clinical centers in Europe has been established. EDD2020 is a crowd-source initiative and provides an evaluation and benchmarking platform for non-clinical researchers to try their novel deep learning based methods and hence promoting research for robust and transferable healthcare systems. A more comprehensive report will be drafted after this challenge.

 The aim of EDD challenge and the organizers is to curate and extend this dataset in future releases through more collaborations and identifying better metrics that can capture generalization of the methods. 
\section*{Acknowledgment}
We would like to acknowledge NIHR Oxford Biomedical Research Centre for funding the  EDD2020 challenge and the workshop. We would also like to thank IEEE International Symposium on Biomedical Imaging (ISBI), 2020 for providing us the opportunity to hold the workshop in conjunction with their conference. Finally, we would like to thank all the participants for submitting their work to this challenge.
\bibliography{endoCV2020.bib}{}

\begin{thebibliography}{1}

\bibitem{pascal-voc-2012}
M.~Everingham, L.~Van~Gool, C.~K.~I. Williams, J.~Winn, and A.~Zisserman.
\newblock The {PASCAL} {V}isual {O}bject {C}lasses {C}hallenge 2012 {(VOC2012)}
  {R}esults.

\bibitem{dutta2019vgg}
Abhishek Dutta and Andrew Zisserman.
\newblock {The VGG} image annotator ({VIA}).
\newblock {\em arXiv preprint arXiv:1904.10699}, 2019.

\bibitem{ALI2020:SREP}
Sharib Ali, Felix Zhou, Barbara Braden, Adam Bailey, Suhui Yang, Guanju Cheng,
  Pengyi Zhang, Xiaoqiong Li, Maxime Kayser, Roger~D. Soberanis-Mukul, Shadi
  Albarqouni, Xiaokang Wang, Chunqing Wang, Seiryo Watanabe, Ilkay Oksuz,
  Qingtian Ning, Shufan Yang, Mohammad~Azam Khan, Xiaohong~W. Gao, Stefano
  Realdon, Maxim Loshchenov, Julia~A. Schnabel, James~E. East, Geroges
  Wagnieres, Victor~B. Loschenov, Enrico Grisan, Christian Daul, Walter
  Blondel, and Jens Rittscher.
\newblock An objective comparison of detection and segmentation algorithms for
  artefacts in clinical endoscopy.
\newblock {\em Scientific Reports}, 10, 2020.

\end{thebibliography}
\bibliographystyle{unsrt}

\end{document}